# Dosimetry study of high repetition rate MeV electron beam from a continuous-wave photocathode gun

Jianhan Sun[1], Jianfeng Lv[1, 2], Shang Tian[1], Juntao Liu[1], Zihao Zhang[1, 2], Hang Xu[1], Lin Lin[1], Senlin Huang[1 a)]

[1]*State Key Laboratory of Nuclear Physics and Technology and Institute of Heavy Ion Physics, School of Physics, Peking University, Beijing 100871, China*

[2] *Beijing Laser Acceleration Innovation Center, Beijing, 101407, China*

a)Authors to whom correspondence should be addressed: huangsl@pku.edu.cn

## Abstract

DC-SRF-II gun, a high-brightness continuous-wave photocathode gun, has greater potential in electron beam irradiation applications. This paper presents the in-vacuum and in-air irradiation dosimetry study of the high repetition rate electron beam from the DC-SRF-II gun with both Monte Carlo simulations and experiments. Especially, high-dose uniform irradiations with flexible and accurate tuning of dose rate across orders of magnitude are demonstrated. Good stability and repeatability of the doses are also shown. The ultra-wide tuning range and precise control of irradiation time and dose rate are expected to pave the way for innovative applications across a wide range of fields.

## 1. Introduction

Electron beam irradiation has been widely applied in various fields such as biomedicine and materials science[1]. In biomedicine, electron beam irradiation can be used for radiotherapy[2,3]. Recently, electron beams have shown potential advantages in ultrahigh dose rate radiotherapy (FLASH-RT), a new paradigm that deliver doses with the dose rate larger than 40 Gy/s[4] [5,6]. In materials science, electron beam irradiation offers an efficient and gentle way[7,8] to modify the structure and properties of materials. For instance, it can be used to regulate surface defects in perovskite, thereby enhancing the quality of perovskite thin films[9]; it can also be used to create NV color centers in diamonds[10–12] and 4H-SiC so as to prepare solid-state quantum bits[13–15], quantum precision measurement magnetometers[10,11,16], and photoluminescent devices[17].

Superconducting radio-frequency (SRF) photocathode guns, as an advanced electron source, offer greater potential to enhance the application of electron beam irradiations. Due to the extremely low surface resistance of the accelerating cavity, SRF guns can operate in continuous wave (CW) mode, enabling the generation of high repetition rate (up to GHz), high-brightness, megavolt (MeV) electron beams[18]. While the electron beam is produced through the photoelectric effect, the bunch charge, current, and temporal profile of electron beam can be flexibly adjusted by manipulating the photocathode drive laser. This allows an ultra-wide tuning range and a precise control of irradiation time and dose rate, opening up new possibilities for applications in various fields including FLASH-RT[19] and material irradiation.

This study investigates the dosimetry characteristics of the high repetition rate electron beam from a CW photocathode gun. Especially, we study the dose delivery of electron beams with both Monte Carlo simulations and irradiation experiments. For the experiments, we have established an irradiation platform utilizing the DC-SRF-II gun, a high-brightness hybrid photocathode SRF gun at Peking University[20]. Both external beam irradiation and in-vacuum irradiation have been carried out as part of the study.

## 2. Methods and Material

### 2.1 The electron gun and beamline

The irradiation experiment utilized the electron beam from the DC-SRF-II photocathode gun. An overall layout of the DC-SRF-II gun, the transport beamline, and the experimental system is depicted in Fig. 1(a). The electron beam produced by the DC-SRF-II gun has an energy of approximately 2 MeV. It travels through the transport beamline and arrives at the experimental station, where the electron beam energy can be measured by a 90° dipole magnet, while its temporal profile and current can be measured by a fast current transformer (FCT) and a Faraday cup, respectively.

The experimental station comprises two sample chambers, as illustrated in Fig. 1(b). One chamber operates in a vacuum environment at a pressure level of $10^{-7}$ Pa, while the other is in an air environment. The beam pipes connecting these chambers are separated by a 0.25 mm thick beryllium window, sealed with a rubber ring gasket. During in-vacuum irradiation, the sample's rotational and vertically translational movements are managed by a magnetic stick, whose position is controlled by a stepper motor. In the case of in-air irradiation, the scattered beam from the beryllium window passes through a detachable collimator to achieve uniformity. Throughout the experiments, a CCD camera is employed to monitor the target, so as to ensure a safe and stable irradiation of the designated samples.

The temporal profile of the electron beam can be flexibly adjusted by regulating the drive laser. As shown in Fig. 1(c), the electron beam can comprise pulse trains, also known as macro pulses, with a tunable duration ranging from approximately 10 ps (i.e., single pulse) to infinity (i.e., continuous mode) at a step length of 12.3 ns or 1 μs, depending on the interval between the pulses (also referred to as micro pulses or electron bunches)[18,21,22]. The repetition rate/frequency of the macro pulse can be adjusted, and the electron beam bunch charge can be varied from 0.1 fC to 0.1 nC. Additionally, by adjusting the strength of the solenoids in the transport beamline, the focusing of the electron beam can be regulated with a beam size at mm or sub-mm level.

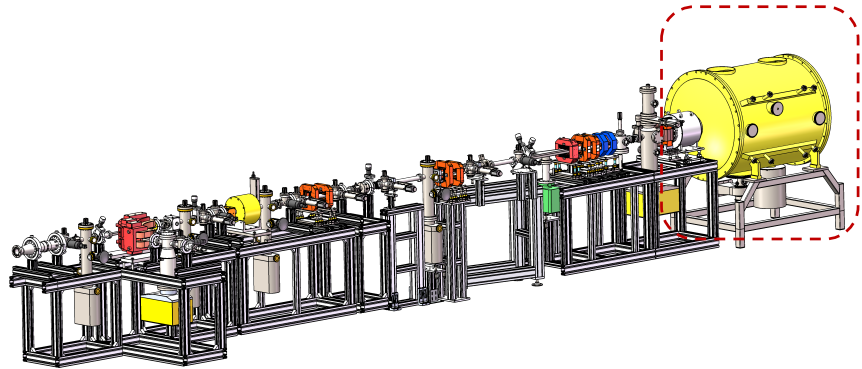

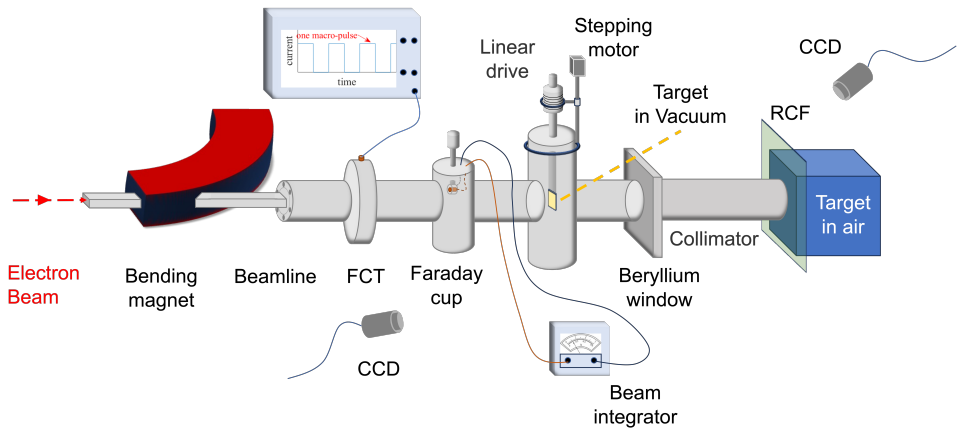

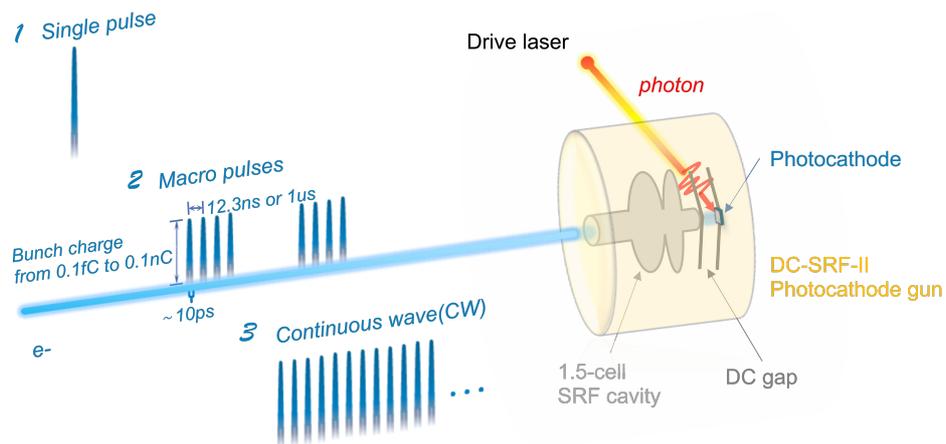

Figure 1 (a) Layout of the DC-SRF-II photocathode gun and beamline. (b) Diagram of the irradiation experimental setup. (c) Electron beam temporal profiles.

**2.2 Simulation setup: TOPAS model configuration and utilized packages**

The simulation of dose depositions of electron beams on radiochromic films (RCFs) and targets is conducted using TOPAS (TOol for PArticle Simulation), a Monte Carlo (MC) simulation platform based on Geant4. This platform is specifically designed to simulate the interaction of particles with materials[23] and offers a diverse range of physics lists for various research objectives. In this study, a specific set of physics lists, including g4em-standard_opt4, g4h-elastic_HP, g4decay, g4stopping, g4ion-binarycascade, and g4h-phy_QGSP_BIC_HP, is utilized to simulate the electron beams' generation, propagation, and interaction with materials. The simulation parameters are configured according to the specifications of the DC-SRF-II gun.

In MC simulation tasks, an electron beam consisting of $2 \times 10^6$ monoenergetic (2 MeV) electrons is generated, with a transverse diameter of 2 mm. For the in-vacuum case, this electron beam is injected into a diamond or 4H-SiC phantom with dimensions of $1 \times 1 \times 0.4$ cm$^3$ to evaluate the energy deposition under irradiation. The phantom is divided into $200 \times 200 \times 200$ voxels, enabling a dose spatial distribution resolution of 0.025 μm$^3$. Besides, the simulation assumes the pressure in the vacuum chamber is on the order of $10^{-7}$ Pa, consistent with that in the experiments. For the in-air case, a water phantom with dimensions of $10 \times 10 \times 10$ cm$^3$, providing a dose spatial distribution resolution of 0.125 mm$^3$.

**2.3 Dose measurement method**

Dose measurements are conducted using RCFs, which have a rapid and direct response to radiation dose, while the optical density is insensitive to dose rate[24,25]. These characteristics make RCFs highly suitable for radiation dosimetry studies across a ultra-wide range of dose rates[26]. In our experiments, GAFCHROMIC EBT3 films were utilized, capable of measuring doses within the range from 0.2 Gy to 100 Gy[24]. The optical density of the film was analyzed using a color flatbed scanner, Epson Expression 10000XL (Seiko Epson Corp., Nagano, Japan), along with the EPSON SCAN software (v3.04)[27,28].

**2.4 Photoluminescence spectrum measurement**

The photoluminescence (PL) spectrum of $V_{Si^-}$ in 4H-SiC was measured using an FLS980 spectrophotometer equipped with an R5509 near-infrared photomultiplier tube (NIR-PMT) as the detector. The excitation wavelength was 785 nm, and an 830 nm long-pass filter was used to eliminate the impact of laser light on the spectrum. For the in-vacuum 4H-SiC irradiation study, experiments were performed at 5 different doses. Due to the orders of magnitude difference in the fluorescence intensity, the samples for the 5 experiments were divided into two groups, each having the same spectrum measurement conditions. Among the 5 samples, the one irradiated with a dose of $10^7$ Gy were measured under both conditions so as to provide a same reference for fluorescence intensity normalization. The measurements of the photoluminescence spectrum were conducted at room temperature.

## 3. Results

**3.1 In-air electron beam irradiation simulation**

First, the dose distribution of in-air irradiation was investigated. The distance between the beryllium window and the target, i.e., the water phantom, was defined as the *delivery distance*, which varied from 1 mm to 245 mm. The simulation model is sketched in Fig. 2(a), and the results, including the dose distribution on the target surface, the dose distribution in the target, the percentage depth dose (PDD), and the integral depth dose (IDD), are shown in Fig. 2(b) to (e). In this study, the simulated doses are normalized by the total charge of the electrons irradiating the phantom, and therefore have a unit of Gray per nanocoulomb (Gy/nC).

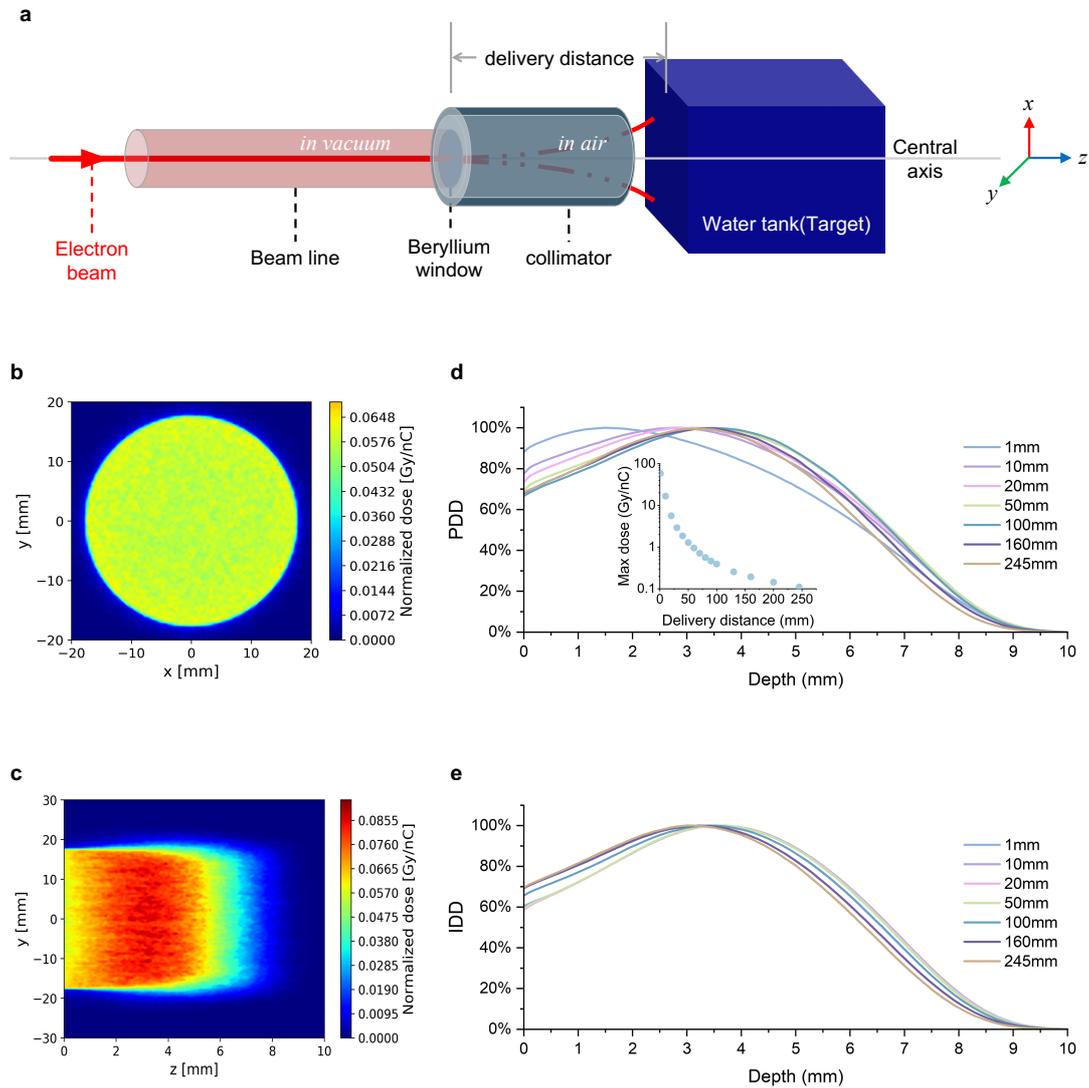

Figure 2 (a) Schematic setup for simulating in-air electron beam irradiation. (b) Dose distribution on the target surface at a delivery distance of 245 mm. (c) Dose distribution in the target at a delivery distance of 245 mm. (d) Percentage depth dose (PDD) at various delivery distances, with an inset displaying the maximum dose versus delivery distances. (e) Integral depth dose (IDD) at different delivery distance. More details can be found in the supplementary materials.

As shown in Fig. 2(b), the dose distribution is uniform at the target surface. This uniformity is a result of the initially homogeneous electron distribution, the large scattering angle of the electrons in the beryllium, and the collimation provided by the collimator between the beryllium window and the target. The dose distribution largely remains uniform inside the target. Although the dose penumbra near the edge broadens with increasing penetration depth, the distinguishable dose distribution area maintains a size close to that of the collimator, as illustrated in Fig. 2(c).

The PDD curves in Fig. 2(d) were obtained by calculating the ratio of the absorbed doses around the central axis ($D_d$) at various depths to its maximum value ($D_{max}$). In contrast, the IDD curves in Fig. 2(e) were calculated as the ratio of the absorbed doses across the cross-sectional extent at each depth to its maximum value. The inset of Fig. 2(d) plots the maximum absorbed dose $D_{max}$ for each delivery distance. It suggests that when the target is placed very close to the beryllium window, e.g., at 1 mm downstream from the window, a maximum dose of above 60 Gy/nC could be achieved. In this case, the dose is essentially distributed in an area with the same size to the electron beam. As the delivery distance increases to 10 mm, the full width at half maximum (FWHM) of the dose spot is about 5 mm and the maximum dose is 17 Gy/nC. When the delivery distance further increases to 245 mm, only the electrons with small scattering angles pass through the collimator, which has an inner diameter of 35 mm. This leads to the uniform dose distributions in Fig. 2(b) and (c), while the maximum dose is reduced to 0.1 Gy/nC.

We also investigated the impact of the beryllium window thickness on the deposited dose distribution in the water phantom. Fig. 3 shows both the longitudinal and transverse dose distributions at various window thicknesses (10 μm, 50 μm, 100 μm, and 250 μm) and delivery distances (1 mm, 10 mm, and 245 mm). At the delivery distance of 1 mm, the dose distribution is less sensitive to the variation of the beryllium window, while at the delivery distance of 10 mm and 245 mm, significant changes can be observed. These demonstrate the effect of the enhanced scattering as the window thickness increases. Analyses of the angular distribution and energy spectrum of the scattered electron beam were also conducted, with the results shown in Fig. 4. As the window thickness increases from 10 μm to 250 μm, the electron energy spectrum shows substantial broadening and the mean energy decreases by about 4.6%. On the other hand, the scattering angle distribution shows more significant broadening and the average scattering angle increases from 1.15° to 8.90°.

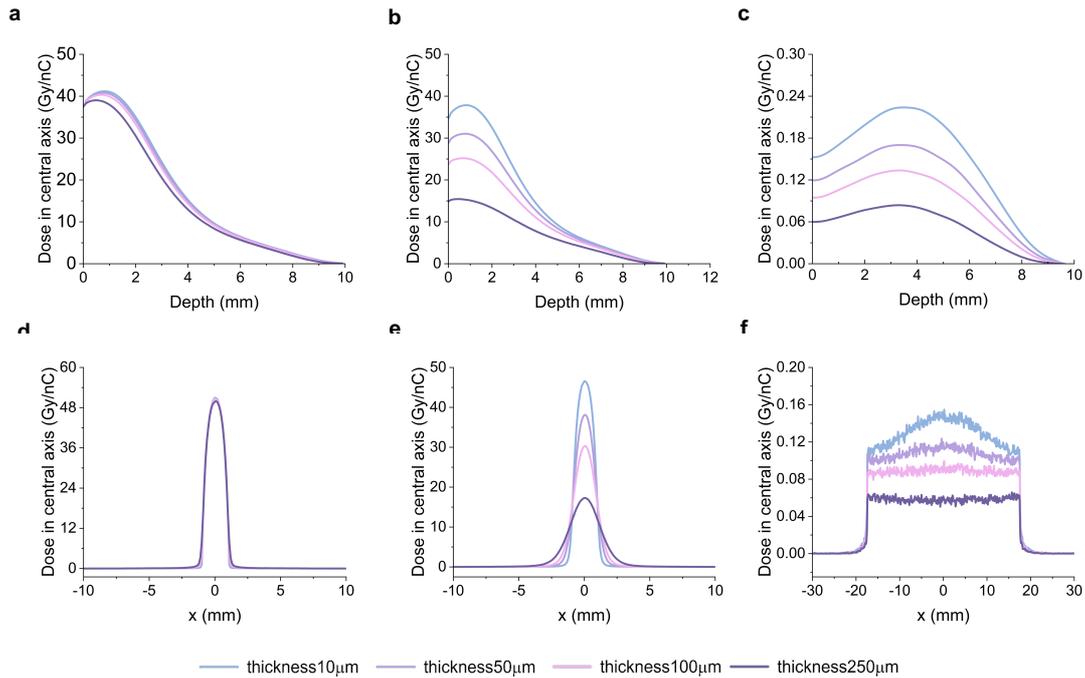

Figure 3 Longitudinal dose distribution (along z axis) in the water phantom at a delivery distance of 1 mm (a), 10 mm (b), and 245 mm (c) and transverse dose distribution (along x axis) at the delivery distance of 1 mm (d), 10 mm (e), and 245 mm (f).

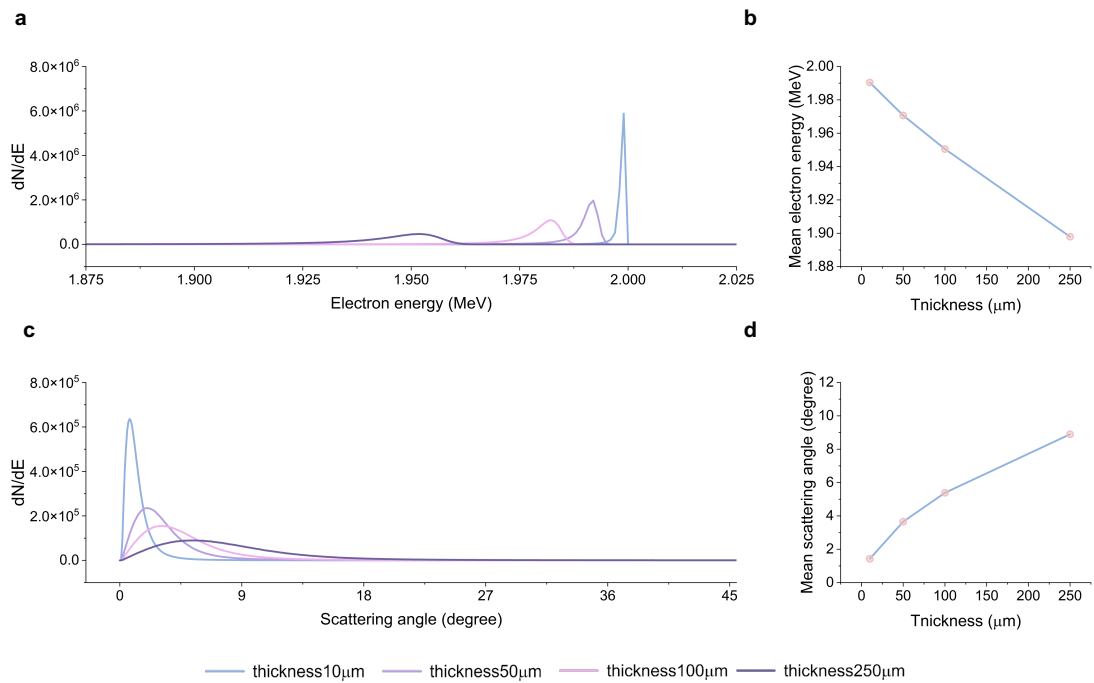

Figure 4 Energy and scattering angle of electrons passing through the beryllium window. (a) energy spectrum; (b) mean energy of the electrons versus beryllium window thickness; (c) scattering angular distribution; (d) mean scattering angle versus beryllium window thickness.

## 3.2 In-air electron beam irradiation experiments

Experiments were conducted to characterize the in-air electron beam irradiation doses using the RCFs. In the experiment, the average current of the electron beam was set to 4.25 nA, 725 nA, 7.25 µA, 63 µA, and 425 µA, corresponding to an averaged surface dose rate of 0.36 Gy/s, 60 Gy/s, 600 Gy/s, 5270 Gy/s, and 36550 Gy/s, respectively. Fig. 5(a) shows the measured doses versus the total charge of the incident electrons along with the Monte Carlo simulation results, while Fig. 5(b) shows the average dose rates versus the electron beam current. Here the average dose rates were calculated by dividing the doses by the irradiation time of the RCFs, while the beam current equals to the total electron charge divided by the time . The average dose rates have a linear dependency on the beam current as $D\ [\text{Gy s}^{-1}] = 0.0684 I\ [\text{nA}]$ . Compared to the Monte Carlo simulation result $D\ [\text{Gy s}^{-1}] = 0.0699 I\ [\text{nA}]$, the difference between the slopes is less than 3%.

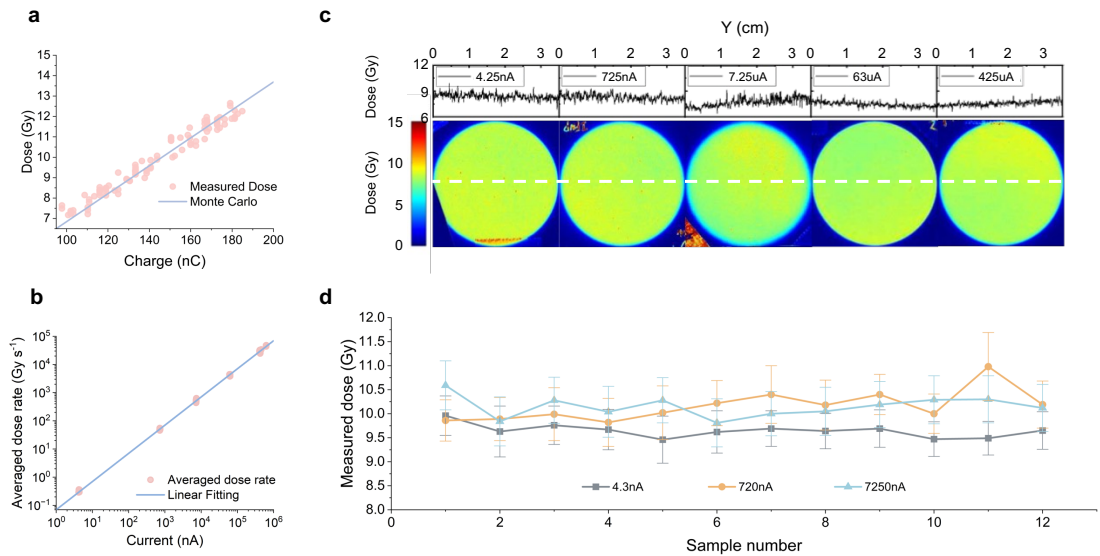

Figure 5 Measurement results of in-air electron beam irradiation doses. (a) The measured doses (using RCFs) versus the total charge of the incident electrons and Monte Carlo simulation results. (b) Average dose rate versus electron beam currents and linear fitting result. (c) Transverse dose distributions at various dose rates (with a same preset dose of 8.7 Gy). (d) Repeated dose measurement results.

Fig. 5(c) displays the transverse dose distribution measured by the RCFs. For all the measurements, the preset doses were kept the same at 8.7 Gy. Note the doses in our experiments were preset by choosing proper electron beam current and time duration according to the linear equation fitted above. The differences between the measurement results and the preset values are less than 5%, showing a good consistency.

The dose stability and repeatability were evaluated, too. Fig. 5(d) shows the dose measurement results at the electron beam currents of 4.3 nA, 720 nA, and 7.25 μA. The measurement at each current was repeated by 10 times with a preset dose of 10 Gy. The deviation of the measured doses from the preset value was less than 5% for all measurements, demonstrating a good stability and repeatability.

### 3.3 In-vacuum electron beam irradiation dosimetry

Electron beam irradiation in vacuum is widely used for defect creation in materials. While 4H-SiC and diamond are important materials for the preparation and study of structural defects like color centers, we select them as samples to study the dose distribution of the in-vacuum irradiation. Fig. 6(a) illustrates the setup for in-vacuum electron beam irradiation simulation, where the electron beam is vertically incident onto a 400 μm thick target. As depicted in Fig. 6(b), in both 4H-SiC and diamond samples, the dose distributions are uniform within the area of the electron beam, which add up to 94.4% and 95.8% of the total doses, respectively. Although the dose penumbra in the vicinity of the edge broadens with the increase of penetration depth, the distributions are still close to a plateau. The doses reach a maximum at 352 μm and 368 μm in the 4H-SiC and diamond samples, respectively, while the average doses are 48.54 Gy/nC and 48.51 Gy/nC, respectively. These uniformly distributed high doses are beneficial for rapid preparation of abundant homogeneous defects in the samples.

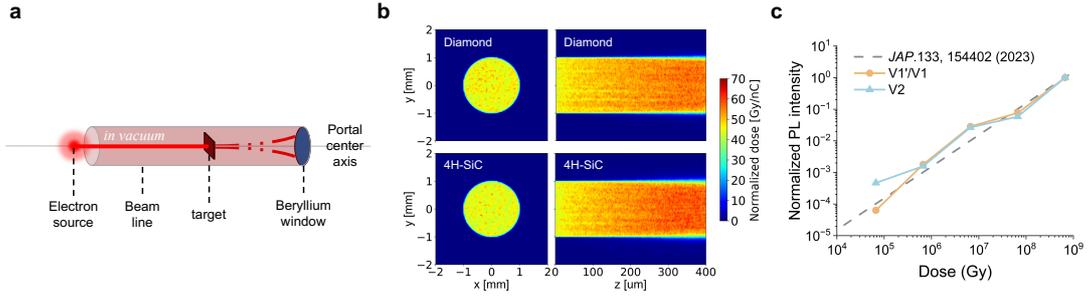

Figure 6 (a) Schematic setup for in-vacuum electron beam irradiation study. (b) Simulated transverse dose distribution (left) and longitudinal dose deposition (right) in diamond and 4H-SiC. (c) Normalized photoluminescence (PL) intensity at room temperature versus irradiation dose (experimental result) along with the exponential fitting result.

The accuracy of the simulation can be verified by characterizing the normalized photoluminescence (PL) intensity of the irradiated samples. For such a purpose, in-vacuum 4H-SiC irradiation experiments were conducted with the electron fluence ranging from $1\times10^{14}$ cm$^{-2}$ to $1\times10^{18}$ cm$^{-2}$. This corresponds to an irradiation dose from $6.7\times10^4$ Gy to $6.7\times10^8$ Gy. The measured PL intensity shows a good exponential dependency on the irradiation dose with a fitted exponent of 1.00±0.10, as illustrated in Fig. 6(c). The result is close to Motoki's work[29] at low irradiation fluxes, where the fitted exponent was 1.02±0.07.

## 4. Discussion

Irradiation with electron beam from high repetition rate photocathode guns exhibits wide range, flexible tuning of dose rate. With a 2 mm diameter, 2 MeV energy electron beam from the DC-SRF-II gun, the in-air irradiation dose conversion efficiency can reach 60 Gy/nC (see the inset of Fig. 2(d)), corresponding to an averaged dose rate of $6\times10^7$ Gy/s at 1 mA electron beam current. Even after the scattered electrons were flattened by a small opening angle collimator (with a length of 245 mm and an inner diameter of 35 mm), the dose rate can still reach $5.9\times10^4$ Gy/s at 1 mA. In this study, we have experimentally demonstrated the tuning of dose rates from $10^{-1}$ Gy/s to $10^4$ Gy/s (see Fig. 5(a)), across 5 orders of magnitude. Our recent experiments have shown that the DC-SRF-II gun has the capability of operating at 3 mA or even higher beam current[30]. Meanwhile, by reducing the bunch charge and duty cycle of the electron beam, the lower limit of electron beam current can be extended to picoampere (pA) level[31,32]. This allows a flexible and accurate tuning of dose rate over 9 orders of magnitude.

A wide tuning range of dose rates allow various experiments with one single device. This may open up new possibilities for radiobiology researches, such as the mechanism of FLASH-RT. In the in-air experiments conducted with the DC-SRF-II gun, radiation biological effects at different dose rates from 0.36 to 36550 Gy/s has been investigated[19]. On the other hand, the in-vacuum irradiation has reached a dose rate at $10^7$ Gy/s level with a uniform transverse distribution and tunable temporal profile. This provides a powerful tool for preparing special structures and investigating the defect formation mechanism in materials. Besides, in the field of radiochemistry, the significance of dose rate is being recognized. Studies have shown that, even under a same total dose, the value of dose rate could have important impacts on the conversion rate[33–35] and the microstructure of products[36].

## 5. Conclusions

We have investigated the in-vacuum and in-air irradiation dosimetry characteristics of the widely tunable high repetition rate electron beam from the DC-SRF-II gun with Monte Carlo simulations and irradiation experiments. For both the in-vacuum and in-air irradiation cases, the results agree well, enabling the enhancement of experimental findings through simulations. We have shown the capability of achieving uniformly distributed high doses with the DC-SRF-II electron beam. We have demonstrated the tuning of in-air irradiation dose rates from $10^{-1}$ Gy/s to $10^4$ Gy/s, across 5 orders of magnitude, and the flexible tuning of in-vacuum irradiation dose from $6.7\times10^4$ Gy to $6.7\times10^8$ Gy. We have shown the good consistency between the preset doses and the measurement results and also demonstrated a good stability and repeatability of the doses. These performances display the great advantages of CW photocathode guns in irradiation applications. In the near future, a flexible and accurate tuning of dose rate across 9 orders of magnitude would be achieved with the DC-SRF-II gun. This ultra-wide tuning range and precise control of irradiation time and dose rate are expected to open up new possibilities for applications across a wide range of fields.

## Supplementary Material

See the supplementary material for the changes of dose distribution under different delivery distance of the high repetition rate beams that transport in the water.

## Acknowledgments

This work is supported by the National Key Re- search and Development Program of China (Grant No. 2016YFA0401904 and 2017YFA0701001) and the State Key Laboratory of Nuclear Physics and Technology, Peking University (Grant No. NPT2022ZZ01).

## Author declarations

### Conflict of Interest

The authors have no conflicts to disclose.

### Author Contributions

**Jianhan Sun:** Conceptualization (lead), Formal Analysis(equal), Investigation(equal), Software(equal), Resources (equal), Visualization(lead), Writing/Original Draft Preparation(lead), Writing/Review & Editing(equal). **Jianfeng Lv:** Data Curation(lead), Formal Analysis(equal), Investigation(equal), Methodology(equal), Resources (equal), Software(equal), Writing/Review & Editing(equal). **Shang Tian:** Investigation(equal), Validation(lead). **Juntao Liu:** Investigation(equal), Resources (equal). Zihao Zhang: Investigation(equal), Writing/Review & Editing(equal). **Lin lin:** Investigation(equal), Supervision(equal). **Hang Xu:** Investigation(equal), Resources (equal). **Senlin Huang:** Funding Acquisition(lead), Project Administration(lead), Resources (equal) , Supervision(equal), Writing/Review & Editing(equal).

## Data Availability

The data that support the findings of this study are available from the corresponding authors upon reasonable request.

## Reference

[1] S. Möller, *Accelerator Technology: Applications in Science, Medicine, and Industry* (Springer International Publishing, Cham, 2020).

[2] J. Lv, X. Zhao, J. Liu, D. Wu, G. Yang, M. Kang, and X. Yan, "Dose rate assessment of spot-scanning very high energy electrons radiotherapy driven by laser plasma acceleration," J. Appl. Phys. **133**(19), 194901 (2023).

[3] G. Sgouros, L. Bodei, M.R. McDevitt, and J.R. Nedrow, "Radiopharmaceutical therapy in cancer: clinical advances and challenges," Nat. Rev. Drug Discov. **19**(9), 589–608 (2020).

[4] V. Favaudon, L. Caplier, V. Monceau, F. Pouzoulet, M. Sayarath, C. Fouillade, M.-F. Poupon, I. Brito, P. Hupé, J. Bourhis, J. Hall, J.-J. Fontaine, and M.-C. Vozenin, "Ultrahigh dose-rate FLASH irradiation increases the differential response between normal and tumor tissue in mice," Sci. Transl. Med. **6**(245), 245ra93-245ra93 (2014).

[5] Y. Ma, Z. Zhao, W. Zhang, J. Lv, J. Chen, X. Yan, X. Lin, J. Zhang, B. Wang, S. Gao, J. Xiao, and G. Yang, "Current Views on Mechanisms of the FLASH Effect in Cancer Radiotherapy," (2024).

[6] M.-C. Vozenin, J. Bourhis, and M. Durante, "Towards clinical translation of FLASH radiotherapy," Nat. Rev. Clin. Oncol. **19**(12), 791–803 (2022).

[7] B. Campbell, and A. Mainwood, "Radiation Damage of Diamond by Electron and Gamma Irradiation," Phys. Status Solidi A **181**(1), 99–107 (2000).

[8] C. Kasper, D. Klenkert, Z. Shang, D. Simin, A. Gottscholl, A. Sperlich, H. Kraus, C. Schneider, S. Zhou, M. Trupke, W. Kada, T. Ohshima, V. Dyakonov, and G.V. Astakhov, "Influence of Irradiation on Defect Spin Coherence in Silicon Carbide," Phys. Rev. Appl. **13**(4), 044054 (2020).

[9] B. Jin, D. Zhao, F. Liang, L. Liu, D. Liu, P. Wang, and M. Qiu, "Electron-Beam Irradiation Induced Regulation of Surface Defects in Lead Halide Perovskite Thin Films," Research **2021**, (2021).

[10] H. Zheng, J. Xu, G.Z. Iwata, T. Lenz, J. Michl, B. Yavkin, K. Nakamura, H. Sumiya, T. Ohshima, J. Isoya, J. Wrachtrup, A. Wickenbrock, and D. Budker, "Zero-Field Magnetometry Based on Nitrogen-Vacancy Ensembles in Diamond," Phys. Rev. Appl. **11**(6), 064068 (2019).

[11] H. Zheng, Z. Sun, G. Chatzidrosos, C. Zhang, K. Nakamura, H. Sumiya, T. Ohshima, J. Isoya, J. Wrachtrup, A. Wickenbrock, and D. Budker, "Microwave-Free Vector Magnetometry with Nitrogen-Vacancy Centers along a Single Axis in Diamond," Phys. Rev. Appl. **13**(4), 044023 (2020).

[12] F.A. Hahl, L. Lindner, X. Vidal, T. Luo, T. Ohshima, S. Onoda, S. Ishii, A.M. Zaitsev, M. Capelli, B.C. Gibson, A.D. Greentree, and J. Jeske, "Magnetic-field-dependent stimulated emission from nitrogen-vacancy centers in diamond," Sci. Adv. **8**(22), eabn7192 (2022).

[13] S.-W. Jeon, J. Lee, H. Jung, S.-W. Han, Y.-W. Cho, Y.-S. Kim, H.-T. Lim, Y. Kim, M. Niethammer, W.C. Lim, J. Song, S. Onoda, T. Ohshima, R. Reuter, A. Denisenko, J. Wrachtrup, and S.-Y. Lee, "Bright Nitrogen-Vacancy Centers in Diamond Inverted Nanocones," ACS Photonics **7**(10), 2739–2747 (2020).

[14] C.P. Anderson, E.O. Glen, C. Zeledon, A. Bourassa, Y. Jin, Y. Zhu, C. Vorwerk, A.L. Crook, H. Abe, J. Ul-Hassan, T. Ohshima, N.T. Son, G. Galli, and D.D. Awschalom, "Five-second coherence of a single spin with single-shot readout in silicon carbide," Sci. Adv. **8**(5), eabm5912 (2022).

[15] K.C. Miao, J.P. Blanton, C.P. Anderson, A. Bourassa, A.L. Crook, G. Wolfowicz, H. Abe, T. Ohshima, and D.D. Awschalom, "Universal coherence protection in a solid-state spin qubit," Science **369**(6510), 1493–1497 (2020).

[16] A. Kuwahata, T. Kitaizumi, K. Saichi, T. Sato, R. Igarashi, T. Ohshima, Y. Masuyama, T. Iwasaki, M. Hatano, F. Jelezko, M. Kusakabe, T. Yatsui, and M. Sekino, "Magnetometer with nitrogen-vacancy center in a bulk diamond for detecting magnetic nanoparticles in biomedical applications," Sci. Rep. **10**(1), 2483 (2020).

[17] V. Ivády, H. Zheng, A. Wickenbrock, L. Bougas, G. Chatzidrosos, K. Nakamura, H. Sumiya, T. Ohshima, J. Isoya, D. Budker, I.A. Abrikosov, and A. Gali, "Photoluminescence at the ground-state level anticrossing of the nitrogen-vacancy center in diamond: A comprehensive study," Phys. Rev. B **103**(3), 035307 (2021).

[18] X. Wen, Study of High Repetition Rate THz Superradiant Radiation Based on SRF Accelerators, D, Peking University, 2016.

[19] J. Lv, J. Sun, Y. Luo, J. Liu, D. Wu, Y. Fang, H. Lan, L. Diao, Y. Ma, Y. Li, M. Wang, Z. Zhao, H. Wang, A. Morris, W. Zhang, Z. Zhang, L. Lin, H. Jia, C. Wang, T. Li, G. Mourou, S. Huang, G. Yang, and X. Yan, "FLASH Irradiation Regulates IFN-β induction by mtDNA via Cytochrome c Leakage," 2024.04.10.588811 (2024).

[20] S. Huang, K. Liu, K. Zhao, and J. Chen, "DC-SRF photocathode gun," Chin. Sci. Bull. **68**(9), 1036–1046 (2022).

[21] T. Wang, H. Xu, Z. Liu, X. Zhang, J. Liu, J. Xu, L. Feng, J. Li, K. Liu, and S. Huang, "Advanced drive laser system for a high-brightness continuous-wave photocathode electron gun," Opt. Express **32**(6), 9699–9709 (2024).

[22] F. Liu, Research on Arbitrary Spatiotemporal Shaping of Photocathode Drive Lasers, D, Peking University, 2020.

[23] J. Perl, J. Shin, J. Schümann, B. Faddegon, and H. Paganetti, "TOPAS: An innovative proton Monte Carlo platform for research and clinical applications," Med. Phys. **39**(11), 6818–6837 (2012).

[24] A. Niroomand-Rad, S.-T. Chiu-Tsao, M.P. Grams, D.F. Lewis, C.G. Soares, L.J. Van Battum, I.J. Das, S. Trichter, M.W. Kissick, G. Massillon-JL, P.E. Alvarez, and M.F. Chan, "Report of AAPM Task Group 235 Radiochromic Film Dosimetry: An Update to TG-55," Med. Phys. **47**(12), 5986–6025 (2020).

[25] S. Devic, "Radiochromic film dosimetry: Past, present, and future," Phys. Med. **27**(3), 122–134 (2011).

[26] P. Casolaro, L. Campajola, G. Breglio, S. Buontempo, M. Consales, A. Cusano, A. Cutolo, F. Di Capua, F. Fienga, and P. Vaiano, "Real-time dosimetry with radiochromic films," Sci. Rep. **9**(1), 5307 (2019).

[27] L. Campajola, P. Casolaro, and F.D. Capua, "Absolute dose calibration of EBT3 Gafchromic films," J. Instrum. **12**(08), P08015 (2017).

[28] S. Devic, N. Tomic, and D. Lewis, "Reference radiochromic film dosimetry: Review of technical aspects," Phys. Med. **32**(4), 541–556 (2016).

[29] S. Motoki, S. Sato, S. Saiki, Y. Masuyama, Y. Yamazaki, T. Ohshima, K. Murata, H. Tsuchida, and Y. Hijikata, "Optically detected magnetic resonance of silicon vacancies in 4H-SiC at elevated temperatures toward magnetic sensing under harsh environments," J. Appl. Phys. **133**(15), 154402 (2023).

[30] H. Jia, T. Li, T. Wang, Y. Zhao, X. Zhang, H. Xu, Z. Liu, J. Liu, L. Lin, H. Xie, L. Feng, F. Wang, F. Zhu, J. Hao, S. Quan, K. Liu, and S. Huang, "High Performance Operation of a Direct-Current and Superconducting Radio-Frequency Combined Photocathode Gun," (2024).

[31] F. Liu, S. Huang, S. Si, G. Zhao, K. Liu, and S. Zhang, "Generation of picosecond pulses with variable temporal profiles and linear polarization by coherent pulse stacking in a birefringent crystal shaper," Opt. Express **27**(2), 1467–1478 (2019).

[32] F. Liu, S. Huang, K. Liu, and S. Zhang, "Highly Stable Linearly Polarized Arbitrary Temporal Shaping of Picosecond Laser Pulses," (JACOW Publishing, Geneva, Switzerland, 2019), pp. 3682–3685.

[33] "Removal of dibenzothiophene from simulated petroleum by integrated γ-irradiation and Zr/alumina catalyst," Appl. Catal. B Environ. **71**(1–2), 108–115 (2007).

[34] Z. Qu, N. Yan, Y. Zhao, J. Jia, and D. Wu, "Removal of Dibenzothiophene from the Simulated Petroleum by γ-Irradiation Induced Reaction," Energy Fuels **20**(1), 142–147 (2006).

[35] N.-Q. Yan, Z. Qu, J.-P. Jia, X.-P. Wang, and D. Wu, "Removal Characteristics of Gaseous Sulfur-Containing Compounds by Pulsed Corona Plasma," Ind. Eng. Chem. Res. **45**(19), 6420–6427 (2006).

[36] C. Hu, L. Cheng, L. Zhou, Z. Jiang, P. Gan, S. Cao, Q. Li, C. Chen, Y. Wang, M. Mostafavi, S. Wang, and J. Ma, "Radiolytic Water Splitting Sensitized by Nanoscale Metal–Organic Frameworks," J. Am. Chem. Soc. **145**(9), 5578–5588 (2023).